# Practical 1-μm GHz fiber comb on silica-based platform

Ruoao Yang[1,5,*], Xingang Jin[2], Ya Wang[3], Minghe Zhao[1,4], Zhendong Chen[1], Xinpeng Lin[2], Fei Meng[1], Duo Pan[1], Qian Li[4], Jingbiao Chen[1], Aimin Wang[1], Zhigang Zhang[1]

1. State Key Laboratory of Photonics and Communications, School of Electronics, Peking University, Beijing, 100871, China
2. Jiaxing Xurui Electronics Tech Co Ltd, Jiaxing, 314001, China
3. State Key Laboratory of Information Photonics and Optical Communications, Beijing University of Posts and Telecommunications, Beijing 100876, China
4. The School of Electronic and Computer Engineering, Peking University, Shenzhen, Guangdong 518055, China
5. State Key Laboratory of Nuclear Physics and Technology, and Key Laboratory of HEDP of the Ministry of Education, CAPT, School of Physics, Peking University, Beijing 100871, People's Republic of China

* ruoao.yang@pku.edu.cn

## Abstract

We present a fully stabilized 1-GHz Yb-fiber laser frequency comb built on silica substrates, utilizing "optical cubes" to house all optical components, ensuring long-term stability and practical operation. Both the femtosecond laser and *f*-to-2*f* interferometer are constructed to silica bricks, with a compact footprint of 290 mm × 250 mm, and a total weight of 1.8 kg. This system provides a stable repetition rate, offset frequency, and a supercontinuum spanning 460-1560 nm without requiring amplification. The carrier-envelop offset frequency exhibits exceptional in-loop stability, with a fractional frequency instability of 3.07 $\times 10^{-18}$ at a 1 second averaging time, improving to 2.12 × $10^{-20}$ at a 10,000 second, maintaining uninterrupted operation for over 60 hours. This work demonstrates a high-performance GHz fiber-based frequency comb, paving the way for applications beyond laboratory environments, including dual-comb spectroscopy, astronomical spectrograph calibration, and portable optical clocks.

## Introduction

Optical frequency combs (OFCs) have revolutionized many fields in modern optics and physics by establishing a direct and phase-coherent connection between the optical and microwave spectral domains[1, 2]. This has enabled them finding extensive use in fields such as the next generation of optical atomic clocks[3,4], accurate broadband spectroscopy[5, 6], ultralow noise microwave extraction[7], precise time-frequency transfer [8], calibration of astronomical spectrographs[9, 10], and quantum optics [11]. Recently, their superior characteristics have been successfully applied in the realization of nuclear clocks[12].

Long-term performance of a comb is crucial in many impactful research areas. For example, calibrating astronomical spectrographs to detect Earth-like exoplanets requires frequency combs that operate reliably for months to years[13]. Meanwhile, the emergence of transportable optical clocks demands compact and robust OFCs capable of faithfully transferring their long-term stability and enabling inter-clock comparisons [14].

Depending on the application scenarios, the comb line spacing ranges from tens of kilohertz to hundreds of gigahertz. In particular, GHz repetition rates offer a compelling balance—facilitating high-speed dual-comb spectroscopy[15], three-dimensional imaging[16], and seamless interfacing with electronics, and nonlinear broadening schemes[17].

Efforts to develop self-referenced GHz OFCs have spanned gain media across the spectrum: titanium at 0.8 μm[18, 19], ytterbium at 1 μm[20-25], erbium at 1.56 μm[17, 26, 27], and chromium at 2.35 μm[28]. Among these, Ytterbium-based systems are especially attractive for their near-unity quantum efficiency[29, 30] and short operating wavelength.

Yet, most GHz Yb combs still rely on solid-state mode-locked lasers with Kerr-lens or semiconductor saturable absorber mirror (SESAM) [24, 25]. Although fiber-based combs are making headway at GHz regime, achieving robust, broadband, and fully stabilized systems remains a challenge. Free-space coupling and multi-mode diode pumping introduce noise and instability, and few systems offer verified long-term carrier-envelope offset ( $f_{\mathrm{CEO}}$ ) locking. In particular, conventional nonlinear polarization evolution (NPE) -based GHz fiber laser comb, typically with hybrid free-space layouts, have yet to demonstrate sustained, fully stabilized operation[31].

To address these limitations, we present a fully stabilized 1-μm GHz fiber frequency comb constructed entirely on silica substrates. The system integrates a femtosecond laser oscillator, supercontinuum generation, and f-to-2f self-referencing interferometer into a compact, rigid, and fully fiber-coupled platform with a footprint comparable to A4 paper. This comb delivers an octave-spanning spectrum (460–1560 nm) without requiring amplification or compression, and supports long-term locking of both the repetition rate and carrier-envelope offset frequency over multiple days.

With this innovative architecture, we achieved notable in-loop frequency stability, reaching a fractional Allan deviation of $3.07\times10^{-18}$ at a 1-second, and $2.12\times10^{-20}$ at a 10,000-second averaging time, scaled to an optical wavelength of 532 nm (~563 THz). To the best of our knowledge, the represents the first reported demonstration of such long-term in-loop stability at the 10,000-second level for any GHz-spaced optical frequency comb.

These results highlight a low-cost, compact, and highly stable GHz comb solution, demonstrating that complex optical frequency comb systems can be successfully miniaturized without compromising performance. This advancement paves the way for practical deployment of GHz fiber combs in field applications such as transportable optical clocks, astronomical spectrograph calibration, and dual-comb spectroscopy.

# Results

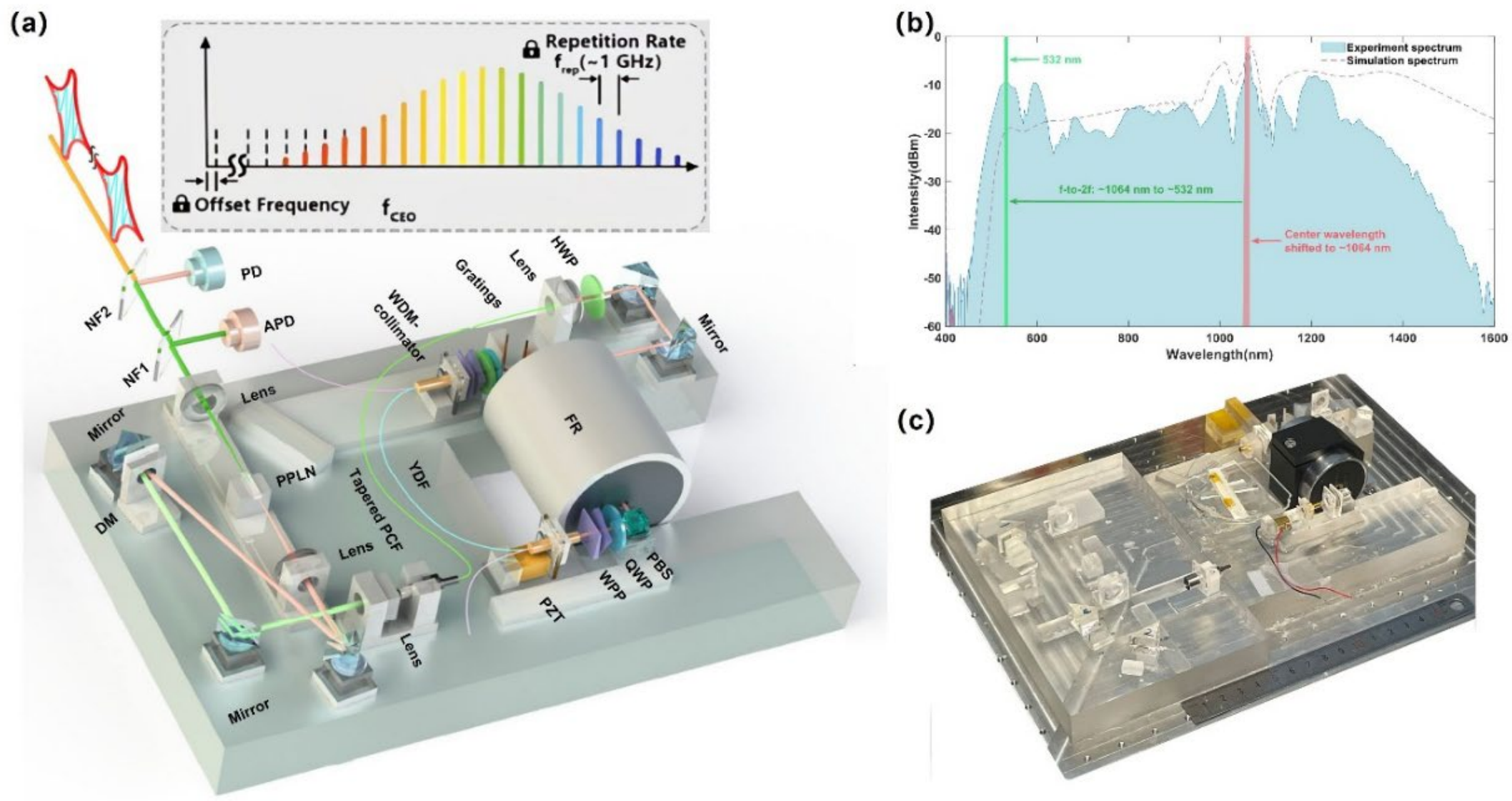

**Figure 1.** (a)Schematic of GHz mode spacing Yb:fiber optical frequency comb, YDF, Yb-doped gain fiber; HWP, half wave plate; FR, Faraday rotator; PBS, polarization beam splitter; QWP, quarter wave plate; WPP, wedge plate pair; PZT, piezo-electric transducer; PPLN, periodically Poled Lithium Niobate; DM, dichroic mirror (Anti-Reflection at 1040 nm, 0° High-Reflection at 532 nm); NF1, notch filter (45° High-Reflection at 532 nm); NF2, notch filter (45° High-Reflection at 1030 nm); PD, photodiode; APD, avalanche photodiode. (b)Measured octave-spanning supercontinuum spectrum generated by a piece of home-made tapered PCF. The dashed purple line represents the simulated output spectrum. (c) Photograph of the GHz Yb:fiber frequency comb on silica bricks.

A comb primarily consists of three parts: a femtosecond pulse oscillator, a supercontinuum generation medium, and a self-referencing interferometer, as shown in Figure 1(a). As the comb source, such femtosecond 1-GHz laser delivers pulses basically the same as our previous work Ref 32. The pulses generated from the 1-GHz laser are directly coupled into a tapered photonic crystal fiber (PCF)[33] for supercontinuum generation. The next to the laser is the offset frequency detection module with standard *f*-to-2*f* technique, which is also integrated in a silica block to ensure the stability.

The RF spectrum of the $f_{\mathrm{CEO}}$ is shown in Fig. 2 (a) and exhibits a signal-to-noise ratio (SNR) of ~44 dB at 300 kHz resolution bandwidth (RBW). Notably, additional beat signals are observed in the RF spectrum at frequencies corresponding to 2 $f_{\mathrm{CEO}}$ and $f_{\mathrm{rep}}$ -2 $f_{\mathrm{CEO}}$ . These spurious signals are attributed to second-harmonic distortion resulting from photodetector saturation.

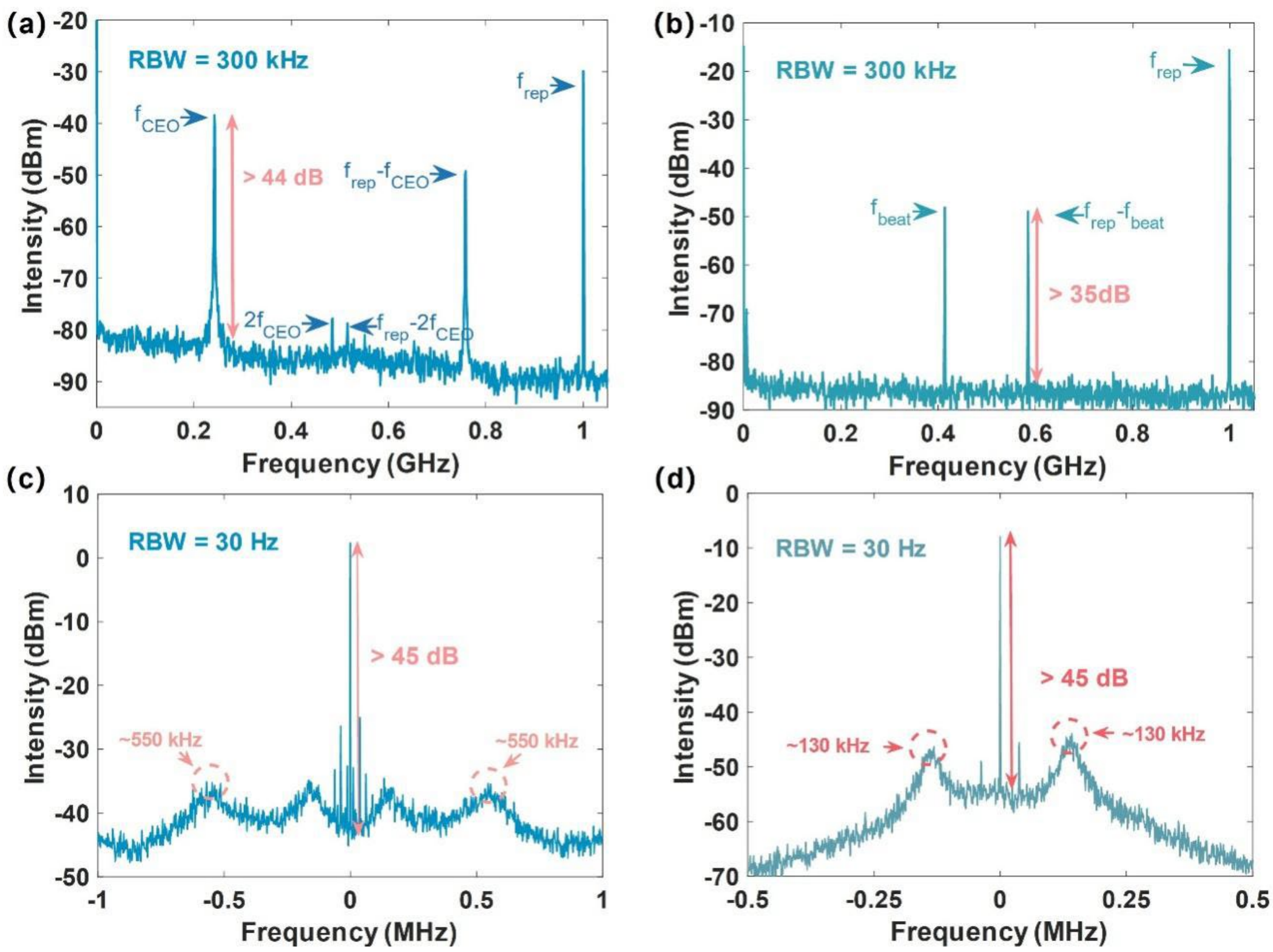

**Figure 2.** (a) RF spectrum of the signal at the output of the *f*-to-2*f* interferometer showing $f_{\mathrm{CEO}}$ , 2 $f_{\mathrm{CEO}}$ , $f_{\mathrm{rep}}$ -2 $f_{\mathrm{CEO}}$ , $f_{\mathrm{rep}}$ - $f_{\mathrm{CEO}}$ and $f_{\mathrm{rep}}$ at 300 kHz RBW. (b) RF spectrum of beat signal between the comb and 1064 nm single-frequency laser, showing $f_{\mathrm{beat}}$ , $f_{\mathrm{rep}}$ - $f_{\mathrm{beat}}$ and $f_{\mathrm{rep}}$ at 300 kHz RBW. (c) RF spectrum of $f_{\mathrm{CEO}}$ locked at 30 Hz RBW. (d) RF spectrum of $f_{\mathrm{beat}}$ locked at 30 Hz RBW.

Meanwhile, in Fig. 2(b), the beat signal between the comb and a single-frequency laser at 1064 nm (Connect VLSP-1064-M-SF) with a linewidth of approximately 50 kHz, shows a beat signal of an SNR greater than 35 dB at 300 kHz RBW, which is sufficiently high for the comb locking to the single-frequency laser. It is worth noting that the linewidth of the beat signal seems much narrower than that of the $f_{\mathrm{CEO}}$ . The broader linewidth of the $f_{\mathrm{CEO}}$ signal is likely due to increased phase noise at the edges of the supercontinuum spectrum, which may be attributed to enhanced relative intensity noise (RIN) introduced during the nonlinear broadening process.

The phase locking of $f_{\mathrm{CEO}}$ and $f_{\mathrm{beat}}$ was achieved by feeding error signals back to the driving current of one of the pump diodes. The locking configuration of the frequency comb is described in the method. The SNRs of both phase-locked signals exceeded 45 dB at an RBW of 30 Hz. The servo bumps at 550 kHz and 130 kHz reflect the bandwidths of the feedback loops for $f_{\mathrm{CEO}}$ and $f_{\mathrm{beat}}$, respectively.

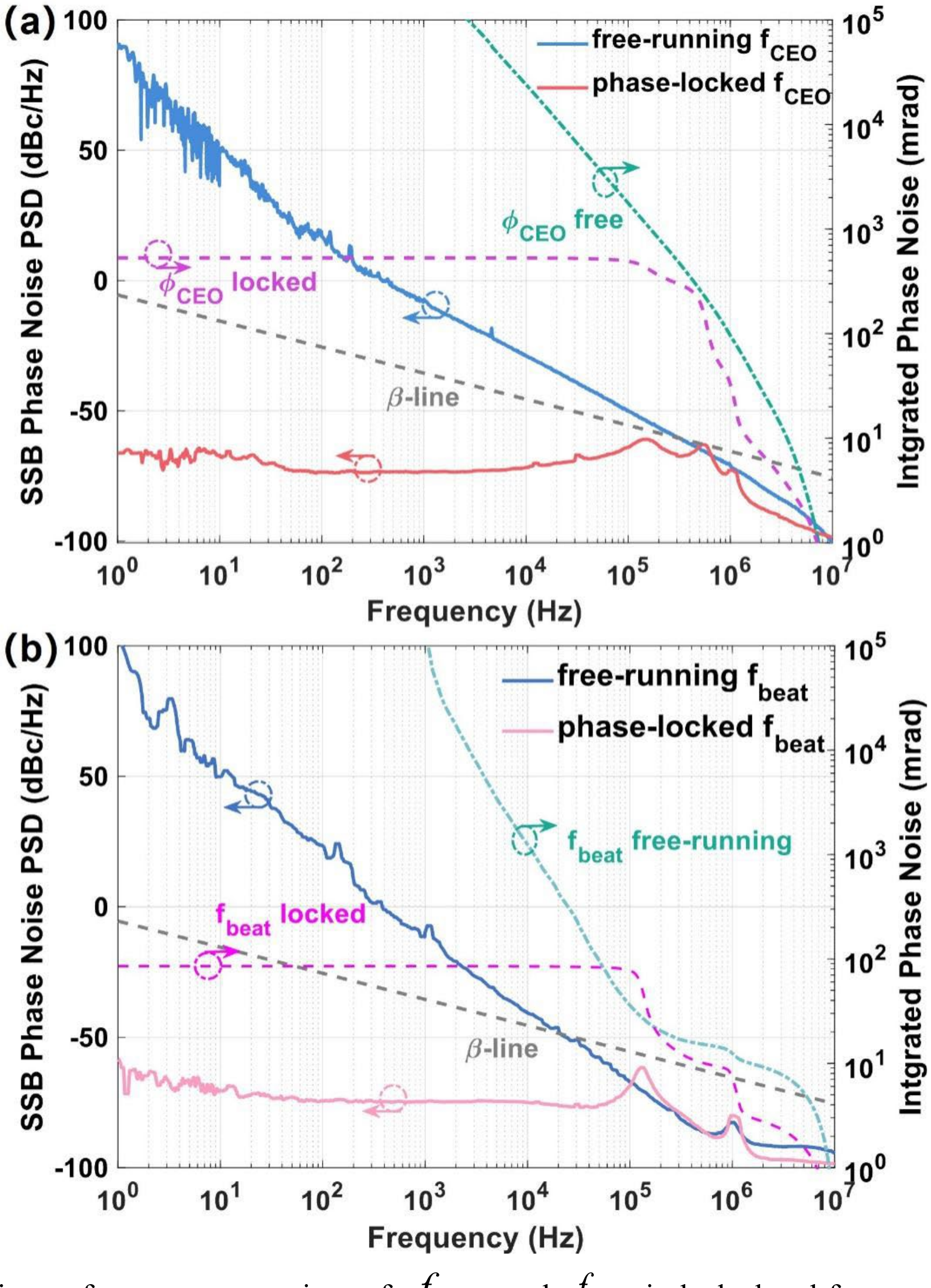

**Figure 3.** Noise performance comparison of $f_{\mathrm{CEO}}$ and $f_{\mathrm{beat}}$ in locked and free-running state. (a) Solid curves: single sideband PN-PSD of the $f_{\mathrm{CEO}}$ in free-running (light blue) and phase-

locked states (red). Dotted: corresponding integrated phase noise from high Fourier frequencies to DC (10 MHz to 1 Hz) in free-running (cyan) and phase-locked states (violet). Gray curve: the $\beta$ -separation line. (b) Solid curves: single sideband PN-PSD of the $f_{\text{beat}}$ in free-running (blue) and phase-locked states (rose), dotted lines: corresponding integrated phase noise in free-running (cyan) and phase-locked states (violet). Gray: the $\beta$ - separation line.

The noise properties of the $f_{\text{CEO}}$ and $f_{\text{beat}}$, both in free-running and stabilized states, were characterized using a phase noise analyzer (Rohde & Schwarz FSWP26). The measured single-sideband phase noise power spectral density (PN-PSD) for both free-running (light blue) and locked (red) $f_{\text{CEO}}$ is shown in Fig.3 (a). The full-span integrated phase noise is 529.41 mrad (from 10 MHz to 1 Hz), primarily contributed by the servo bumps. Two distinct servo bumps are observed in the PN-PSD at 160 kHz and 550 kHz, which align with the positions seen in Fig.2 (c). The residual noise remains well below the $\beta$ -separation line (gray dashed line)[34] across all frequencies, indicating that the carrier-envelope offset frequency does not influence the linewidth of the comb teeth. Fig. 3 (b) shows the measured PN-PSD for both free-running (blue) and locked states (rose) of $f_{\text{beat}}$. The full-span integrated phase noise is 85.57 mrad (from 10 MHz- 1 Hz), and the servo bump observed at 130 kHz corresponds to the one seen in Fig. 2(d). All the RF signals shown in Figures 2 and 3 were measured at the same location as the frequency counter indicated in method Figure 8.

The frequency noise presented above characterizes the high-frequency noise components, whereas the time-domain analysis provides the insights into the long-term behavior of the frequency comb.

To verify the long-term stability, we simultaneously recorded the frequency sequence of $f_{\text{CEO}}$ and $f_{\text{rep}}$ using frequency counters (Keysight 53230A) with 1-second gate time. Fig. 4 (a) and (b) show a continuous 60-hour time series recorded for $f_{\text{CEO}}$ and $f_{\text{rep}}$ with frequency counters. Those resulted in the standard deviation for $f_{\text{CEO}}$ and $f_{\text{rep}}$ of 1.731 mHz and 1.235 mHz, respectively.

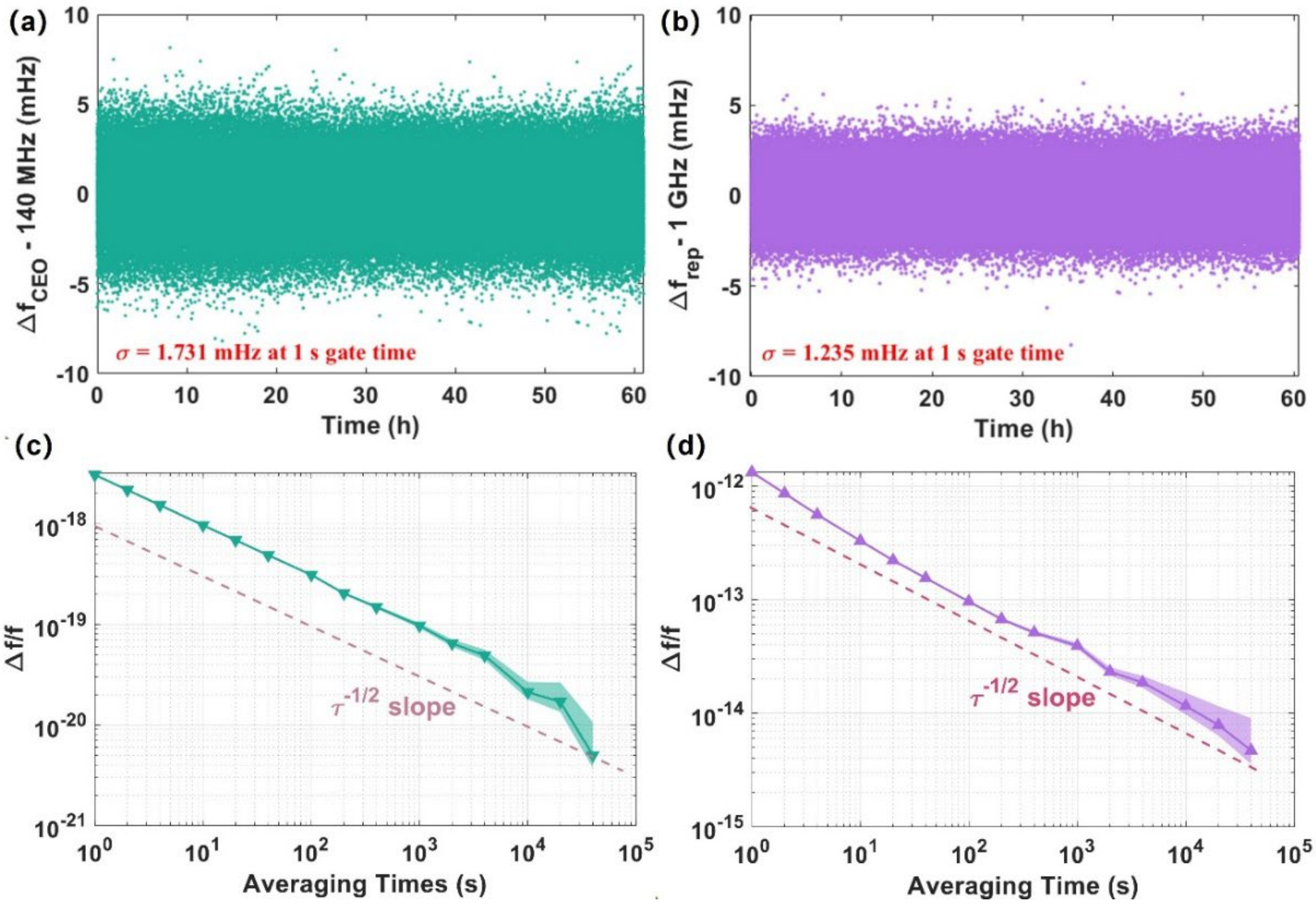

**Figure 4.** (a) Continuous 60-hour recorded time series of the stabilized $f_{\mathrm{CEO}}$, at a fundamental frequency of 140 MHz. (b) Continuous 60-hour recorded time series of the stabilized $f_{\mathrm{rep}}$, with a fundamental frequency of 1 GHz. (c) Allan deviation of $f_{\mathrm{CEO}}$ scaled to an optical frequency. (d) Allan deviation of $f_{\mathrm{rep}}$. The red dash lines in (c) and (d) represent the $\tau^{-1/2}$ dependence due to the dead time of frequency counters. The shaded regions in (c) and (d) represent the error bars.

When scaled to an optical wavelength of 532 nm (~563 THz), which was used for $f_{\mathrm{CEO}}$ detection, the in-loop fractional frequency instability reached $3.07\times10^{-18}$ at a 1-second averaging time, and further decreased to $2.12\times10^{-20}$ at a 10,000-second averaging time, as shown in Fig.4 (c). To the best of our knowledge, this represents the first demonstration of in-loop stability at the 10,000-second level for a GHz-repetition-rate optical frequency comb.

The relative frequency stability of $f_{\mathrm{rep}}$ was measured to be at the level of $1.32\times10^{-12}$ at 1-second averaging time, as shown in Fig.4 (d), which is primarily limited by the reference synthesizer used in the stabilization loop. In both traces, the observed $\tau^{-1/2}$ slope in the log-log plot indicates the presence of white frequency noise, which stems from the dead time of frequency counter, degrading phase coherence and converting white phase noise into white frequency noise.

These results establish a new benchmark for long-term stability in GHz comb systems, demonstrating that our fully fiber integrated platform can sustain high coherence over extended time scales, a critical requirement for applications such as optical clock comparison, astro-comb calibration, and dual-comb spectroscopy.

## Discussion

This work presents a significant advancement in the field of GHz OFCs by demonstrating a fully-locked Yb:fiber laser frequency comb that delivers broadband spectral coverage, low phase noise, and unprecedented long-term in-loop stability.

By integrating silica substrates and holders with a hybrid laser architecture and dispersion-managed PCF, we have developed a complete GHz comb system that maintains notable mechanical and thermal stability. This advanced structure exemplifies the ongoing trend toward miniaturization of optical systems, combining free-space and fiber-based light propagation. The rigid construction significantly reduces drift in the repetition rate. The direct generation of an octave-spanning spectrum, without the need for amplification and compression, ensures efficient $f_{\mathrm{CEO}}$ detection. Although $f_{\mathrm{CEO}}$ is weakly dependent on the cavity length, the use of a short fiber segment reduces environmental sensitivity. However, compared to lower repetition rate combs (100 MHz–250 MHz), GHz combs exhibit a larger $f_{\mathrm{CEO}}$ variation range (0–500 MHz), which adds challenges to the locking circuit. The solidified common-path *f*-to-2*f* interferometer leads to the stability of the $f_{\mathrm{CEO}}$. In addition, the selected low noise diodes play a key role in stabilizing the $f_{\mathrm{CEO}}$.

One of the remarkable features of this comb is the long-term performance. Both $f_{\mathrm{CEO}}$ and $f_{\mathrm{rep}}$ remained continuously locked for more than 60 hours, with the potential for even longer operation. The in-loop Allan deviation of the $f_{\mathrm{CEO}}$ reached $3.07\times10^{-18}$ at 1-second and further decreased to $2.12\times10^{-20}$ at 10,000 seconds, a result that has not been reported for GHz combs before. This demonstrates the robust and stable performance of the GHz comb.

Despite these achievements, certain limitations remain. The pre-positioned cavity components restrict online tuning of the repetition rate, as the PZT translation range is typically insufficient for megahertz-level frequency adjustments. Additionally, the non-polarization-maintaining photonic crystal fiber may introduce additional intensity noise, which will be addressed in future work.

In summary, we have proposed a novel and practical solution for achieving a compact, robust, and cost-effective GHz frequency comb with promising long-term stability. We demonstrated millihertz-level frequency variation and notably low in-loop Allan deviation values of $10^{-18}$ at 1s and $10^{-20}$ at 10 ks. **T**he directly generated octave-spanning supercontinuum (460–1560 nm), coupled with the comb's narrow linewidth, makes it highly suitable for astronomical spectrograph calibration, transportable optical clocks, and other real-world applications. The successful integration of a fiber laser comb on a silica substrate marks a significant step toward the development of GHz frequency combs with potential for deployment beyond laboratory environments.

## Materials and methods

**Architecture of the comb on silica.** The GHz comb is based on a nonlinear polarization rotation (NPR) mode-locking mechanism. To achieve a 1 GHz repetition rate, the laser cavity is minimized in length, suppressing excessive nonlinear phase accumulation and dispersion, which in turn supports the generation of ultrashort pulses. As demonstrated in Ref. 32, all free-space optical components—such as PBSs, gratings, collimators, and waveplate holders—are rigidly bonded onto silica bricks ("optical cubes") and mounted on a silica baseplate. This compact and robust integration ensures turnkey, self-starting operation with excellent thermal and mechanical stability.

The oscillator delivers an average power of 700 mW with only 1.8 W pump power. The pulse spectrum [Fig.5(a)], shows a FWHM of 33 nm centered at 1045 nm. Autocorrelation measurements [Fig. 5(b)] indicate a pulse duration of ~54 fs (assuming a $sech^2$ shape), corresponding to ~0.7 nJ pulse energy and ~13kW peak power. These pulse characteristics are sufficient to directly generate an octave-spanning supercontinuum—eliminating the need for external amplification or compression and keeping total power consumption low (limited to ~ 200 W electrical input for the comb operation, excluding locking electronics).

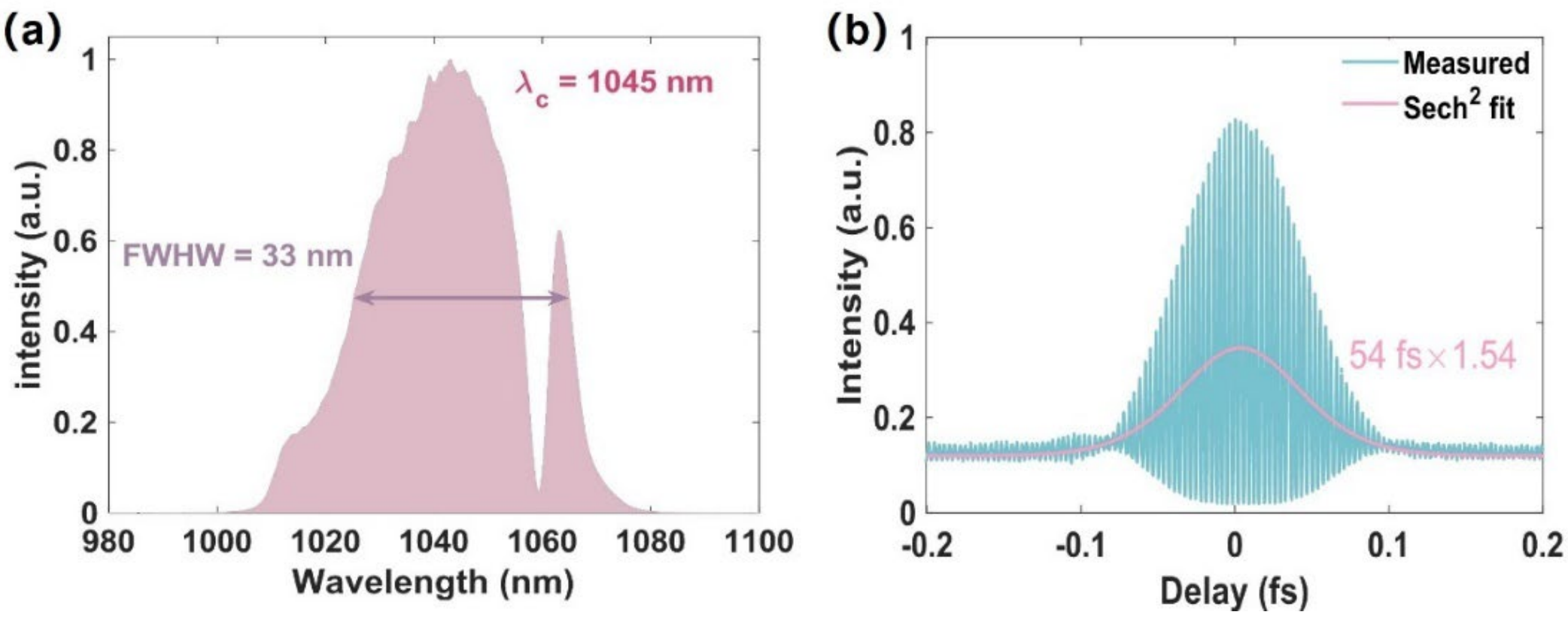

**Figure 5.** (a) Linear spectrum of the GHz "solid-state fiber" laser source. (b) Autocorrelation traces of experimentally measured pulse (cyan) and hyperbolic secant fit (pink).

For $f_{\mathrm{CEO}}$ detection, we employed a solidified, common-path *f*-to-2*f* interferometer. To overcome alignment sensitivity in free-space spatial-dispersion optics, critical components are pre-aligned and permanently fixed with silica-based mounts. The long-wavelength spectral components pass through a dichroic mirror (DM), reflect off a mirror positioned behind the dichroic mirror (DM), and overlap temporally with the short-wavelength light directly reflected by the DM. The overlap gap is tunable to ensure precise temporal alignment. The combined beam is then focused onto a MgO-doped periodically-poled lithium niobate (MgO: PPLN) bulk crystal for frequency doubling. The generated *f* and 2*f* signals are filtered and detected using an avalanche photodiode (Hamamatsu C5658).

To further enhance environmental resilience, the entire system is enclosed within a 360 mm × 255 mm × 90 mm aluminum housing lined with acoustic insulation. The enclosure's baseplate is temperature-stabilized to 25± 0.1 ° C using a thermoelectric cooler. The APD and PD for $f_{CEO}$ and the repetition rate detection are placed outside the enclosure to minimize thermal loading.

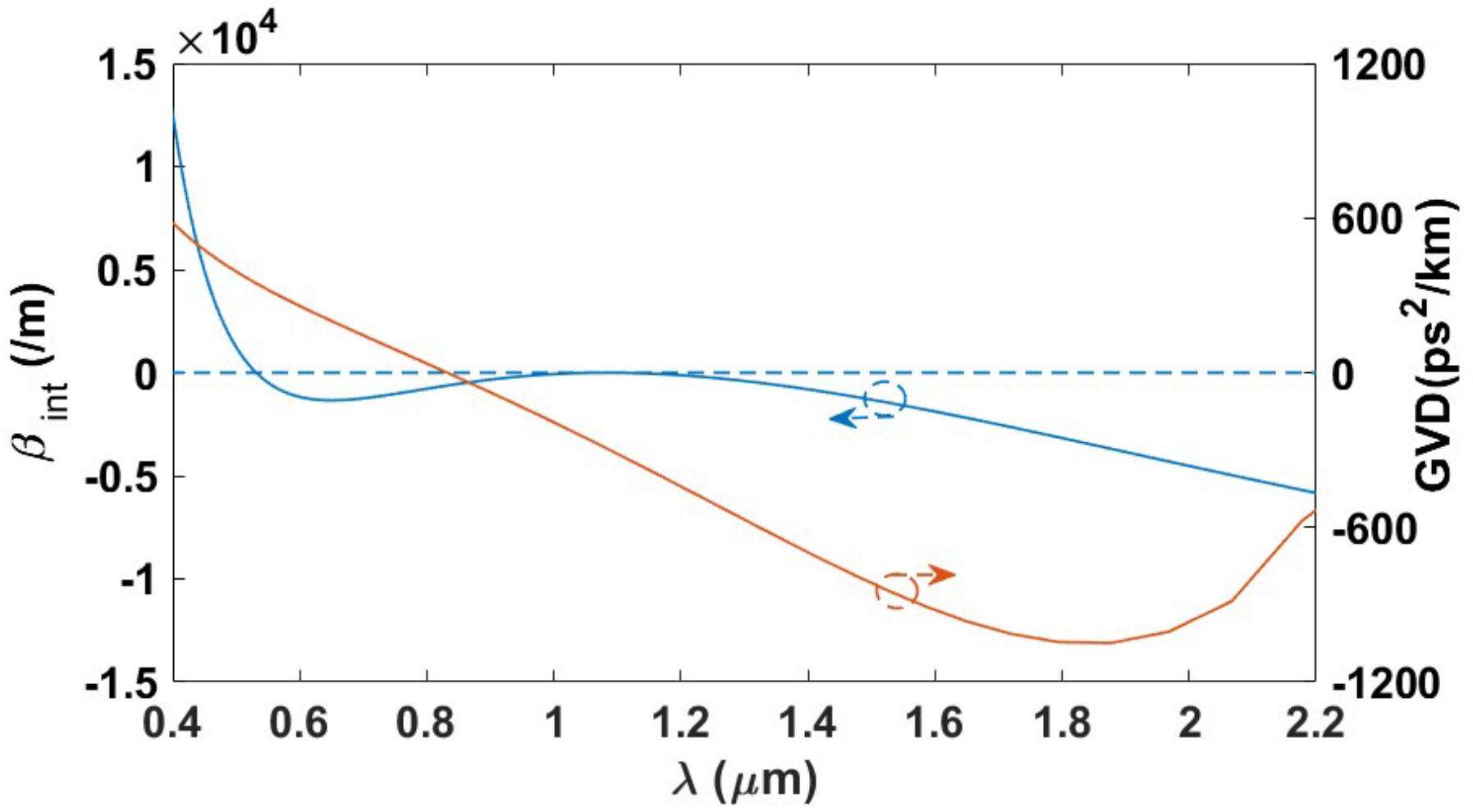

**Figure 6.** The integrated dispersion as a function of the wavelength for the tapered PCF with core diameter of 2.3 μm, pumped at 1045nm (blue). The group velocity dispersion (GVD) for the tapered PCF with core diameter of 2.3 μm.

**Supercontinuum Generation by the tapered photonic crystal fiber.** The tapered photonic crystal fiber (PCF) was specially designed to achieve phase-matching between the pump wavelength at 1045 nm and the dispersive wave generation near 532 nm, which is crucial for $f_{CEO}$ detection. The design approach follows the methodology detailed in Ref. 34. Using Lumerical MODE Solutions, we calculated the PCF's effective index $n_{eff}(\omega)$ and frequency-dependent mode effective area $A_{eff}(\omega)$. These parameters were used to determine the wavelength-dependent propagation constant $\beta_{eff} = n_{eff}(\omega)\cdot\omega / c$, and through differentiation, the dispersion coefficients $\beta_n(\omega)$ and the nonlinear coefficient $\gamma(\omega) = n_{eff}(\omega)\times(\omega+\omega_0)/[cA_{eff}(\omega)]$.

To simulate pulse dynamics, the generalized nonlinear Schrödinger equation (GNLSE) was solved using the fourth-order Runge-Kutta in the interaction picture (RK4IP) method[35,36]. The PCF was tapered from an initial core diameter of 4.1 μm down to 2.3 μm, resulting in a shift of the zero-dispersion wavelength (ZDW) from ~1 μm to ~830 nm. The generation of dispersive waves is determined by the integrated dispersion $\beta_{\text{int}}$, which follows the phase-matching condition[37]:

$$\beta_{\text{int}} = \sum_{n=2}^{\infty} \frac{\beta_n(\omega_{\text{s}})}{n!}(\omega_{\text{d}} - \omega_{\text{s}})^n = \frac{1}{2}\gamma P_{\text{s}} \approx 0 \tag{1}$$

Where $\omega_{\text{s}}$ and $\omega_{\text{d}}$ are the soliton and dispersive wave frequencies, respectively. The integer $n$ represents different dispersion orders, and $P_{\text{s}}$ is the soliton peak power. The term on the right-hand side of Equation. (1) is typically very small and can be neglected. Under these conditions, the integrated dispersion $\beta_{\text{int}}$ becomes zero ~532 nm, as shown in Figure 6. This tapering process not only shifts the ZDW but also enhances the nonlinearity required for efficient supercontinuum generation.

The supercontinuum spectrum was generated by coupling the output pulse into a home-made tapered PCF. As shown in Fig. 1(b), the resulting supercontinuum spans from 460 nm to 1560 nm. The spectral peak shifts from 1045 nm to 1064 nm due to the self-phase modulation (SPM). Notably, a strong dispersive wave emerges near 532 nm, enabling optimal $f_{\text{CEO}}$ detection via *f*-to-2*f* interferometry using an off-the-shelf MgO:PPLN crystal.

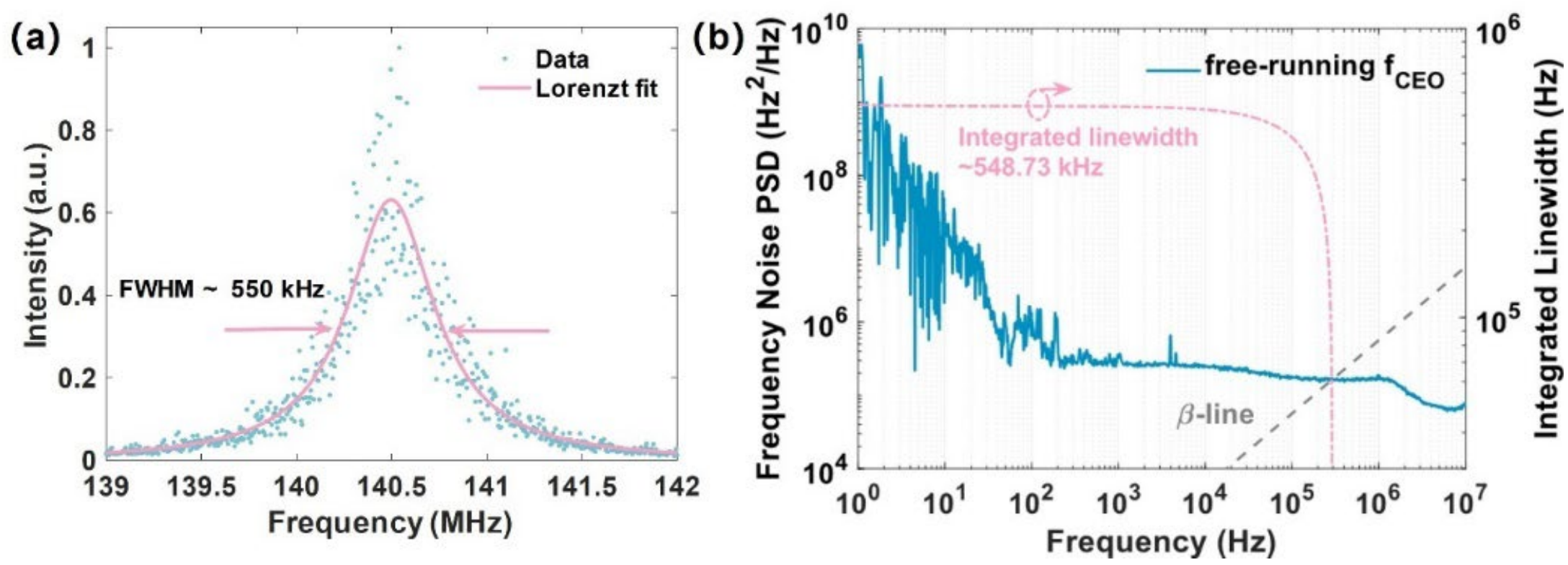

**Figure 7.** (a) The free-running linewidth of $f_{\text{CEO}}$ with Lorentzian fitting is about 550 kHz. (b) Frequency noise power spectral density of the free-running $f_{\text{CEO}}$ (blue) and the corresponding integrated linewidth (pink).

**Phase-locking configuration:** The most concerned of an optical frequency comb is its long-term stability. Before phase locking the $f_{\mathrm{CEO}}$, it is essential to evaluate the linewidth of the $f_{\mathrm{CEO}}$ in free-running conditions. We employed two different methods to perform this evaluation. The first method involves direct observation using a spectrum analyzer (Rigol RSA3030), and the second method calculates the linewidth from the measured frequency noise of the $f_{\mathrm{CEO}}$. By fitting the measured data from the spectrum analyzer with a Lorentzian profile, as shown in Figure 7. (a), the linewidth of $f_{\mathrm{CEO}}$ was approximately 550 kHz. Alternatively, by integrating the frequency noise power spectral density (FN-PSD) from the intersection point with the $\beta$-separation line to low frequencies (the $\beta$-separation line is expressed as $8\ln(2)f/\pi^2$)[34], the linewidth was calculated to be approximately 548.73 kHz (Figure 7. (b)), which is very close to the value obtained using the RF spectral analyzer. This linewidth falls in the bandwidth of the locking electronics.

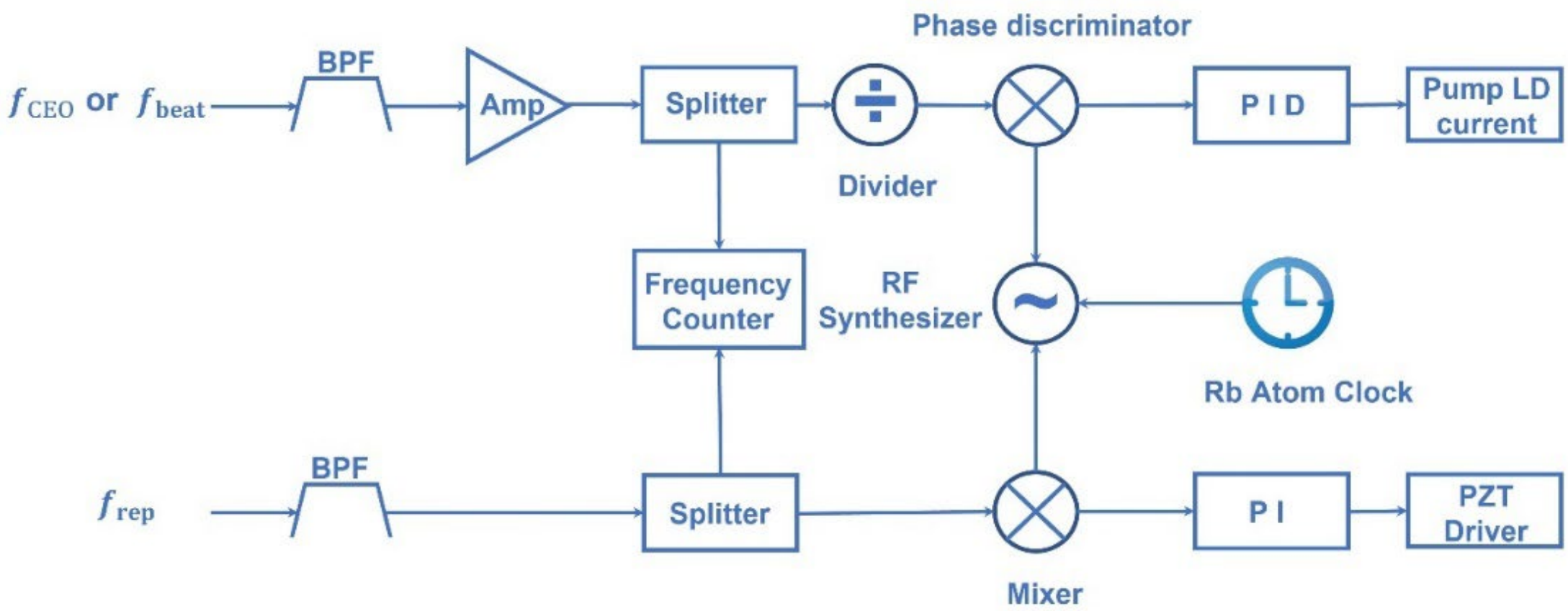

**Figure 8.** Schematic of GHz comb stabilization. BPF: bandpass filter, Amp: amplifier, PID: proportional-integral- derivative controller, PI: proportional-integral controller, LD: laser diode, PZT: piezo-electric transducer.

The phase-locking setup for $f_{\mathrm{CEO}}$ and $f_{\mathrm{beat}}$ is shown in the block diagrams of Figure 8. The $f_{\mathrm{CEO}}$ signal, detected at the output of the *f*-to-2*f* interferometer and set to approximately 140 MHz, was filtered through a bandpass filter and amplified to ~0 dBm. The amplified $f_{\mathrm{CEO}}$ signal was then divided by 8 and compared with a 17.5 MHz reference signal from an external signal synthesizer (Rohde-Schwarz SMA-100A), which was referenced to an Rb clock to ensure long-term stable operation of the frequency comb. The resulting error signal was processed by a PID controller (IMRA Universal Locking Electronics), and the stabilization of the $f_{\text{CEO}}$ was achieved by feeding the feedback signal into the current of one of the pump diodes. The configuration for locking $f_{\text{beat}}$ was similar, but instead of a division factor of 8, a division factor of 4 was used.

The repetition rate ($f_{\text{rep}}$) locking to a RF reference is more straightforward. The fundamental frequency of $f_{\text{rep}}$ was compared in a double-balanced mixer (DBM) to a reference signal from a signal synthesizer (Rigol SMA-100A), which was locked to the same Rb clock used in the $f_{\text{CEO}}$ stabilization. The resulting phase error signal was low-pass filtered (Minicircuit LPF1.9+) and processed by a proportional-integral servo-controller (Newport LB1005). The correction signal generated was then amplified by a high-voltage amplifier (20 × gain) and applied to the PZT used to hold the semi-WDM in the cavity.

The stabilization of $f_{\text{rep}}$ was achieved by feeding back the correction signal to the PZT to control the cavity length. Since the cavity length of the seed laser is inherently very stable (as discussed in Ref. 32), it is easily stabilized using slow PZT modulation.

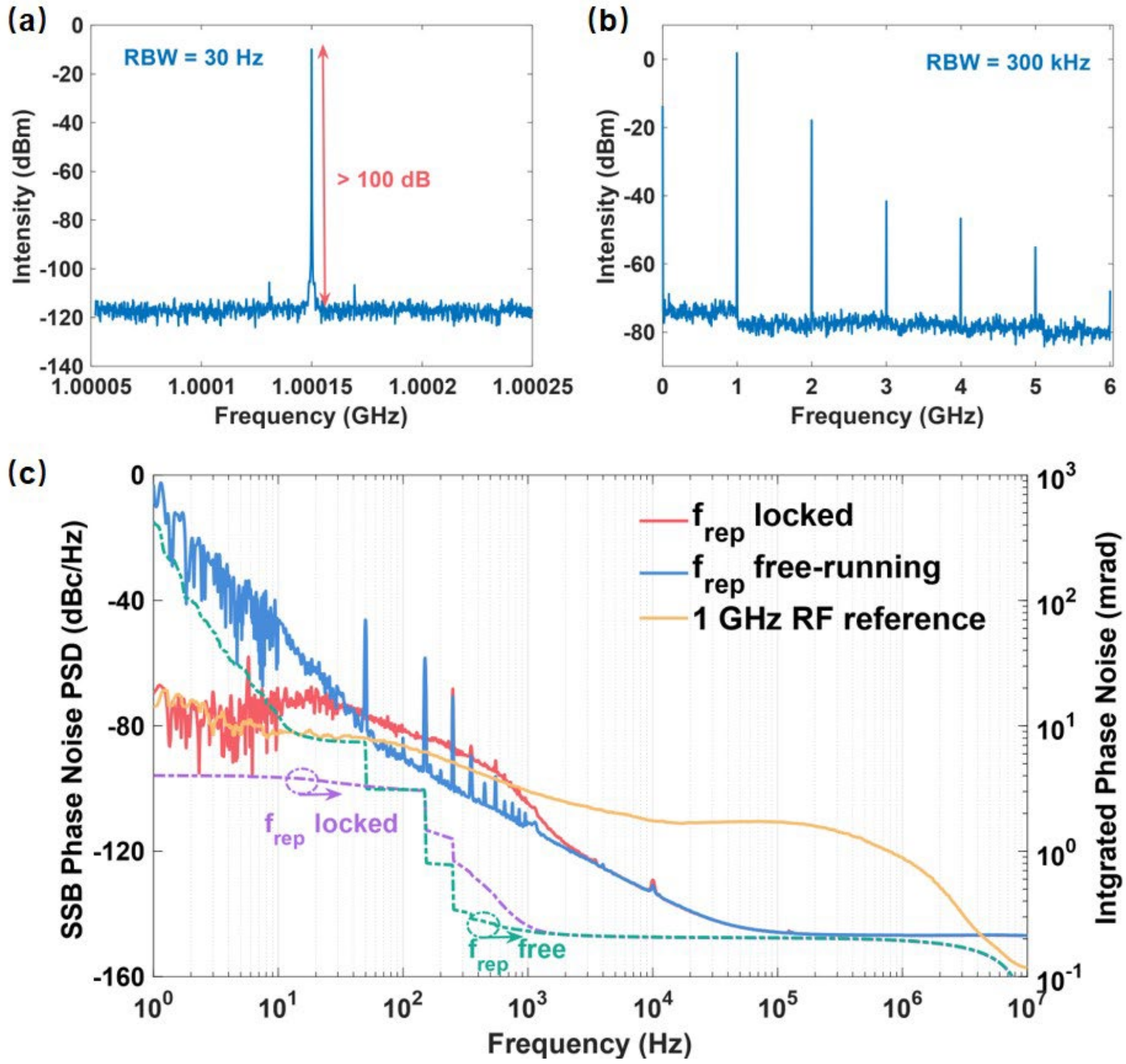

**Figure 9.** (a) RF spectrum around the fundamental repetition rate measured at 30 Hz resolution bandwidth. (b) RF spectrum of higher-order harmonics up to 6 GHz, where the attenuation of higher-order harmonics is limited by the bandwidth of the photodetector used (EOT 3000 A). (c) Noise performance comparison of $f_{\text{rep}}$ in locked and free-running status. Solid curves: single sideband PN-PSD of the $f_{\text{rep}}$ in free-running (blue) and phase-locked status (red). Dotted: corresponding integrated phase noise from high Fourier frequencies to DC (10 MHz to 1 Hz) in free-running (cyan) and phase-locked status (violet). Orange solid: single-sideband PN-PSD of the 1 GHz RF reference signal.

Figure 9. (a) and (b) are the RF spectrum of the repetition rate for two different frequency spans. The SNR exceeding 100 dB (@30 Hz RBW) confirms the stable mode-locking of the 1 GHz pulse train. The resulting PN-PSD of the stabilized $f_{\text{rep}}$ is shown in Figure 9. (c), along with the SSB-PN of the free-running repetition rate. The orange curve represents the PN-PSD of the 1 GHz reference signal (originated from the signal synthesizer, Rohde-Schwarz SMA-100A), which shows a higher phase noise in the range beyond 50 Hz in comparison to the free-running repetition rate signal. Due to the limited servo bandwidth of the PZT, the phase noise of the repetition rate only responded up to 2 kHz, with suppression observed below 50 Hz. However, in the range of 50 Hz to 2 kHz, the phase noise of the stabilized repetition rate was even elevated due to the inferior performance of the reference source compared to that of the optical comb itself.

## Acknowledgements

The authors gratefully acknowledge grant funding from National Natural Science Foundation of China (U2031208), the national Key R&D Program of China (2023YFC3402604). R. Y. acknowledges support from the Boya Postdoctoral Fellowship and Postdoctoral Fellowship Program (GZB20230009).

## Contributions

Z.Z. and R.Y. conceived and designed the experiments. R.Y. performed the simulations and experiments. X. J., X. L. made the "Optical Cubes". Y. W., Z. D. contributed to the experiments. F. M., D. P. and J. C. contributed to the phase-locked technology. M. Z., Q. L. contributed to simulations. A. W. contributed to the supercontinuum generation. R. Y. analyzed the data. Z. Z. supervised the work. R.Y. and Z. Z. wrote the manuscript with input from the others.

## Data availability

The data that support the findings of this study are available from the corresponding author upon reasonable request.

## Conflict of interest

The authors declare no conflicts of interest.

## References

1.S. T. Cundiff and J. Ye. Colloquium: Femtosecond optical frequency combs. *Reviews of Modern Physics* **75**, 325 (2003).

2. Diddams, Scott A., Kerry Vahala, and Thomas Udem. Optical frequency combs: Coherently uniting the electromagnetic spectrum. *Science* **369**, eaay3676 (2020).

3.Rosenband, Till, *et al.* Frequency ratio of Al+ and Hg+ single-ion optical clocks; metrology at the 17th decimal place. *Science* **319**, 1808-1812 (2008).

4.Campbell, Corey J., *et al.* Single-ion nuclear clock for metrology at the 19th decimal place. *Physical review letters* **108**, 120802 (2012).

5.Coddington, Ian, Nathan Newbury, and William Swann. Dual-comb spectroscopy. *Optica* **3**, 414-426 (2016).

6.Picqué, Nathalie, and Theodor W. Hänsch. Frequency comb spectroscopy. *Nature Photonics* **13**, 146-157(2019).

7.Nakamura, Takuma, *et al*. Coherent optical clock down-conversion for microwave frequencies with 10− 18 instability. *Science* **368**, 889-892(2020).

8.Giorgetta, Fabrizio R., *et al*. Optical two-way time and frequency transfer over free space. *Nature Photonics* **7**, 434-438(2013).

9.Li, Chih-Hao, *et al.* A laser frequency comb that enables radial velocity measurements with a precision of 1 cm $s^{-1}$. *Nature* **452**, 610-612 (2008).

10.Steinmetz, Tilo, *et al*. Laser frequency combs for astronomical observations. *Science* **321**, 1335-1337(2008).

11.Sinclair, Laura C., *et al.* Operation of an optically coherent frequency comb outside the metrology lab. *Optics express* **22**, 6996-7006(2014).

12. Zhang, Chuankun, *et al*. Frequency ratio of the 229mTh nuclear isomeric transition and the 87Sr atomic clock. *Nature* **633**, 63-70(2024).

13. Fischer, D. *et al.* State of the field: extreme precision radial velocities. *Publ. Astron. Soc. Pac. **128***, 066001 (2016)

14. Roslund, Jonathan D., *et al*. Optical clocks at sea. *Nature* **628**, 736-740(2024).

15.Hoghooghi, Nazanin, Ryan K. Cole, and Gregory B. Rieker. 11-μs time-resolved, continuous dual-comb spectroscopy with spectrally filtered mode-locked frequency combs. *Applied Physics B* **127**, 17 (2021).

16. Kurata, Shintaro, *et al.* Dead-zone free single-shot three-dimensional measurement using a high-repetition-rate Yb: fiber comb. *Optics Continuum* **1**, 2374-2388 (2022).

17. Lesko, Daniel MB, *et al.* Fully phase-stabilized 1 GHz turnkey frequency comb at 1.56 μm. *OSA Continuum* **3**, 2070-2077(2020).

18. Fortier, Tara M., Albrecht Bartels, and Scott A. Diddams. Octave-spanning Ti: sapphire laser with a repetition rate> 1 GHz for optical frequency measurements and

comparisons. *Optics Letters* **31**, 1011-1013(2006).

19. Chen, Li-Jin, *et al.* Octave-spanning, dual-output 2.166 GHz Ti: sapphire laser. *Optics express* **16**, 20699-20705(2008).

20. Hartl, I., *et al.* Fully stabilized GHz Yb-fiber laser frequency comb. *Advanced Solid-State Photonics*. Optica Publishing Group, 2009

21. Pekarek, Selina, *et al.* Self-referenceable frequency comb from a gigahertz diode-pumped solid-state laser. *Optics express* **19**, 16491-16497 (2011).

22. Endo, Mamoru, Isao Ito, and Yohei Kobayashi. Direct 15-GHz mode-spacing optical frequency comb with a Kerr-lens mode-locked Yb:$Y_2O_3$ ceramic laser. *Optics express* **23**, 1276-1282 (2015).

23. Hakobyan, Sargis, *et al*. Carrier-envelope offset stabilization of a GHz repetition rate femtosecond laser using opto-optical modulation of a SESAM. *Optics Letters* **42**, 4651-4654(2017).

24. Hakobyan, Sargis, *et al*. Full stabilization and characterization of an optical frequency comb from a diode-pumped solid-state laser with GHz repetition rate. *Optics Express* **25**, 20437-20453(2017).

25. Müller, Michael, *et al.* Powerful 1-µm 1-GHz optical frequency comb. *Optics Express* **31**, 44823-44831 (2023).

26 Chao, David, *et al*. Self-referenced Erbium fiber laser frequency comb at a GHz repetition rate. *Optical Fiber Communication Conference*. Optica Publishing Group, 2012.

27. Shoji, Tyko D., *et al*. Ultra-low-noise monolithic mode-locked solid-state laser. *Optica* **3**, 995-998(2016).

28. Smolski, Viktor, *et al*. Half-Watt average power femtosecond source spanning 3–8 µm based on subharmonic generation in GaAs. *Applied Physics B* **124**, 101(2018).

29. Keller, Ursula, and R. Paschotta. Ultrafast lasers. Springer International Publishing, 2021.

30. Okuyucu, Serdar, et al. Diode-pumped passively mode-locked femtosecond Yb: YLF laser at 1.1 GHz. *Optics Express* **32**, 15555-15564 (2024).

31. Li, Chen, *et al*. 1 GHz repetition rate femtosecond Yb: fiber laser for direct generation of carrier-envelope offset frequency. *Applied optics* **54**, 8350-8353 (2015).

32. Yang, Ruoao, *et al*. Attosecond timing jitter from high repetition rate femtosecond "solid-state fiber lasers". *Optica* **9**, 874-877(2022).

33. Yang, Ruoao, *et al*. Flat visible spectrum by a genetic algorithm optimized photonic crystal fiber in the GHz comb spacing. *Optics Letters* **48**, 2829-2832 (2023).

34. Domenico, Gianni D. et al. Simple approach to the relation between laser frequency noise and laser line shape. *Applied optics* **49**, 4801-4807 (2010).

35. Johan Hult, A fourth-order Runge-Kutta in the interaction picture method for

simulating supercontinuum generation in optical fibers. *Journal of Lightwave Technology*, **25**, 3770-3775 (2007)

36. J. Lægsgaard, Mode profile dispersion in the generalized nonlinear Schrödinger equation, *Optics Express*, **15**, 16110-16123 (2007)

37. J. M. Dudley, J. R. Taylor, *Supercontinuum generation in optical fiber*. (Cambridge University Press, 2010).